\documentclass[journal, twoside]{IEEEtran} 
\usepackage{epsf,cite,amsmath,amscd,amssymb,graphicx,latexsym,multicol}
\usepackage{algorithm,algorithmic}
\usepackage{comment}

\newtheorem{proposition}{Proposition}
\newenvironment{thmproof}[1]
{\noindent\hspace{2em}{\it #1 }}
{\hspace*{\fill}~\QED\par\endtrivlist\unskip}

\centerfigcaptionstrue

\begin{document}
\bibliographystyle{ieeetr}

\title{A Utility-Based Approach to Power Control and Receiver Design in Wireless Data Networks}

\author{Farhad~Meshkati,~\IEEEmembership{Student Member,~IEEE,}
H.~Vincent~Poor,~\IEEEmembership{Fellow,~IEEE,} Stuart~C.~Schwartz,~\IEEEmembership{Fellow,~IEEE,}
and~Narayan~B.~Mandayam,~\IEEEmembership{Senior Member,~IEEE}
\thanks{Manuscript received November 12, 2003; revised November 8, 2004 and April 22, 2005.
This research was supported in part by the National Science
Foundation under Grants ANI-03-38807 and CCR-02-05214, and in part
by the New Jersey Center for Wireless Telecommunications. Parts of
this work have been presented at the $37^{th}$ Conference on
Information Sciences and Systems (CISS'03), The Johns Hopkins
University, in March 2003, and at the $41^{st}$ Annual Allerton
Conference on Communication, Control and Computing, University of
Illinois at Urbana-Champaign, in October 2003.}
\thanks{F.~Meshkati, H.~V.~Poor and S.~C.~Schwartz are with
the Department of Electrical Engineering, Princeton University,
Princeton, NJ 08544 USA (e-mail:
{\{meshkati,poor,stuart\}@princeton.edu}). }
\thanks{N.B.~Mandayam is with WINLAB, Rutgers University,
Piscataway, NJ 08854 USA (e-mail: {narayan@winlab.rutgers.edu}).}}

\markboth{IEEE Transactions on Communications,~Vol.~X,
No.~X,~MONTH~X}{Meshkati \MakeLowercase{\textit{et al.}}: A
Utility-Based Approach to Power Control and Receiver Design in
Wireless Data Networks}

\maketitle

\begin{abstract}
In this work, the cross-layer design problem of joint multiuser
detection and power control is studied using a game-theoretic
approach. The uplink of a direct-sequence code division multiple
access (DS-CDMA) data network is considered and a non-cooperative
game is proposed in which users in the network are allowed to
choose their uplink receivers as well as their transmit powers to
maximize their own utilities. The utility function measures the
number of reliable bits transmitted by the user per joule of
energy consumed. Focusing on linear receivers, the Nash
equilibrium for the proposed game is derived. It is shown that the
equilibrium is one where the powers are SIR-balanced with the
minimum mean square error (MMSE) detector as the receiver. In
addition, this framework is used to study power control games for
the matched filter, the decorrelator, and the MMSE detector; and
the receivers' performance is compared in terms of the utilities
achieved at equilibrium (in bits/Joule). The optimal cooperative
solution is also discussed and compared with the non-cooperative
approach. Extensions of the results to the case of multiple
receive antennas are also presented. In addition, an admission
control scheme based on maximizing the total utility in the
network is proposed.
\end{abstract}


\section{Introduction}

Power control is used for resource allocation and interference
management in both the uplink and the downlink of Code Division
Multiple Access (CDMA) systems. In the uplink, the purpose of
power control is for each user to transmit enough power so that it
can achieve the required quality of service (QoS) at the uplink
receiver without causing unnecessary interference to other users
in the system. One approach that has been very successful in
providing insights into design of power control algorithms for
data networks is the game-theoretic approach studied in
\cite{MacKenzieWicker01, GoodmanMandayam00, JiHuang98,
GoodmanMandayam01, Saraydar01, Saraydar02, Xiao01, Zhou01, Alpcan,
Sung, Feng01, MeshkatiMC}. In \cite{MacKenzieWicker01}, the
authors provide motivations for using game theory to study
communication systems, and in particular power control. In
\cite{GoodmanMandayam00} and \cite{JiHuang98}, power control is
modeled as a non-cooperative game in which users choose their
transmit powers in order to maximize their utilities, where
utility is defined as the ratio of throughput to transmit power.
In \cite{GoodmanMandayam01}, a network-assisted power control
scheme is proposed to improve the overall utility of the system.
The authors in \cite{Saraydar01} and \cite{Saraydar02} use pricing
to obtain a more efficient solution for the power control game.
Similar approaches are taken in \cite{Xiao01,Zhou01,Alpcan,Sung}
for different utility functions. Joint network-centric and
user-centric power control is discussed in \cite{Feng01}. In
\cite{MeshkatiMC}, the authors propose a power control game for
multi-carrier CDMA systems. In all the work done so far, the
receiver is assumed to be a simple matched filter. No work has
taken into account the effects of the receiver on power control.
Also, all prior work in this area has concentrated on single
antenna receivers.

This work is the first one that tackles the cross-layer design
problem of joint multiuser detection and power control in the
context of a non-cooperative game-theoretic setting. It attempts
to bring a cross-layer design perspective to all the earlier work
that has studied power control from a game-theoretic point of
view. Our focus throughout this work is on energy efficiency. We
are mainly concerned with applications where it is more important
to maximize the number of bits transmitted per joule of energy
consumed than to maximize throughput. We first propose a
non-cooperative (distributed) game in which users are allowed to
choose their uplink receivers as well as their transmit powers. We
focus on linear receivers and derive the Nash equilibrium for the
proposed game. In addition, we use this framework to study power
control for the matched filter, the decorrelator
\cite{LupasVerdu89} and the minimum mean square error (MMSE)
\cite{MadhowHonig94} receiver. We show that regardless of the type
of receiver, the Nash equilibrium is an SIR-balancing solution
with the same target SIR (signal to interference plus noise ratio)
for all receiver types\footnote{This can also be shown to be true
even for some non-linear receivers \cite{MeshkatiGuo04}.}. Using a
large-system analysis, we derive explicit expressions for the
utilities achieved at equilibrium for each receiver type. This
allows us to compare the performance of these receivers in terms
of energy efficiency (i.e., the number of bits transmitted per
joule of energy consumed). In addition, the performance of our
non-cooperative approach is compared with the optimal cooperative
(centralized) approach. It is shown that only for the matched
filter is the difference in performance between these two
approaches significant. For the decorrelator, the two solutions
are identical, and for the MMSE receiver, the utility achieved by
the Pareto-optimal solution is only slightly greater than that
achieved by the non-cooperative approach. We also extend our
analysis to the case where multiple receive antennas are used at
the uplink receiver. The effects of power pooling and interference
reduction, which are the benefits of using multiple receive
antennas, are demonstrated and quantified for the matched filter,
the decorrelator, and the MMSE receiver in terms of the utilities
achieved at equilibrium. A utility-maximizing admission control
scheme is also presented. We show that using the proposed scheme,
the number of admitted users for the MMSE receiver is greater than
or equal to the total number of admitted users for the matched
filter and decorrelator combined. These results constitute the
first study of power control and receiver design in a unified
framework.

The organization of this paper is as follows. In Section
\ref{background}, we provide the background for this work, while
the system model is given in Section \ref{system model}. We
describe our proposed power control game in Section \ref{NCPCG-SA}
and derive the Nash equilibrium for this game. In Section
\ref{comparison PC-SA}, we use the game-theoretic framework along
with a large-system analysis to compare the performance of various
linear receivers in terms of achieved utilities. The
Pareto-optimal solution to the power control game is discussed in
Section \ref{social optimum} and its performance is compared with
that of the non-cooperative approach. Extensions of our analysis
to multi-antenna systems are given in Section \ref{PC-MA}. We then
present a utility-maximizing admission control scheme in Section
\ref{admission control}. Numerical results and conclusions are
provided in Sections~\ref{numerical
results}~and~\ref{conclusions}, respectively. Throughout this
work, we concentrate on the uplink of a synchronous
direct-sequence CDMA (DS-CDMA) wireless data network.

\section{Background}\label{background}

Consider the uplink of a DS-CDMA data network where each user
wishes to locally and selfishly choose its action in such way as
to maximize its own utility. The strategy chosen by a user affects
the performance of other users in the network through
multiple-access interference. In such a multiple-access network,
there are several questions to ask concerning the interaction
among the users. First of all, what is a reasonable choice of a
utility function that measures energy efficiency? Secondly, given
such a utility function, what strategy should a user choose in
order to maximize its utility? If every user in the network
selfishly and locally picks its utility-maximizing strategy, will
there be a stable state at which no user can unilaterally improve
its utility (Nash equilibrium)? What are some of the properties of
such an equilibrium? How does such a non-cooperative approach
compare with a cooperative scheme?

Game theory is the natural framework for modeling and studying
such an interaction. To pose the power control problem as a
non-cooperative game, we first need to define a utility function
suitable for measuring energy efficiency for data applications.
Most data applications are sensitive to error but tolerant to
delay. It is clear that a higher SIR level at the output of the
receiver will result in a lower bit error rate and hence higher
throughput. However, achieving a high SIR level often requires the
user terminal to transmit at a high power which in turn results in
low battery life. These issues can be quantified (as in
\cite{GoodmanMandayam00}) by defining the utility function of a
user to be the ratio of its throughput to its transmit power,
i.e.,
\begin{equation}\label{eq4a}
    u_k = \frac{T_k}{p_k} \ .
\end{equation}
Throughput here is the net number of information bits that are
transmitted without error per unit time.

Let $f_s(\gamma_k)$ represent the probability that a packet is
received without an error, where $\gamma_k$ is the SIR for user
$k$. Our assumption is that if a packet has one or more bit
errors, it will be retransmitted. Assuming that retransmissions
are independent, the average number of transmissions necessary to
receive a packet correctly is equal to $\frac{1}{f_s(\gamma_k)}$.
Therefore, we have
\begin{equation}\label{eq4}
    T_k = \frac{L}{M} R_k f_s(\gamma_k)  ,
\end{equation}
where $L$ and $M$ are the number of information bits and the total
number of bits in a packet, respectively; and $R_k$ is the
transmission rate for the $k^{th}$ user, which is the ratio of the
bandwidth to the processing gain. The packet success rate (PSR),
which is represented by $f_s (\gamma_k)$, depends on the details
of the data transmission such as modulation, coding, and packet
size. In most practical cases, however, $f_s(\gamma_k)$ is
increasing and has a sigmoidal shape. For example, when the
modulation is BPSK (binary phase shift keying) and the noise is
additive white Gaussian, $f_s(\gamma_k)$ is given by
${\left(1-Q(\sqrt{2\gamma_k}) \right)^M}$, where $Q(\cdot)$ is the
complementary cumulative distribution function of a standard
normal random variable. Notice that, in this case, $f_s(0)=2^{-M}$
is strictly positive due to the possibility of random guessing at
the receiver. This means that based on our definition for the
utility function, a user can potentially achieve infinite utility
by transmitting zero power.

\begin{figure}[t]
\centering
\includegraphics[width=3.5in]{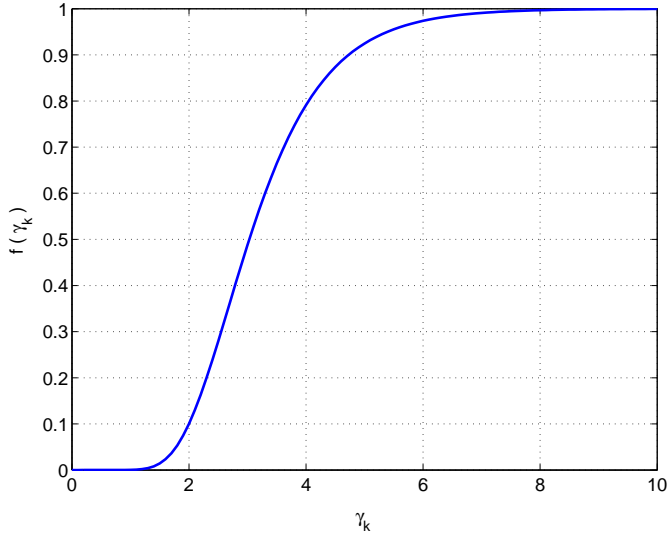}
\caption{A Typical Efficiency Function.} \label{fig1}
\end{figure}

To prevent the above undesirable situation, we replace the PSR
with an \emph{efficiency function}, $f(\gamma_k)$, when
calculating the throughput for our utility function. The
efficiency function should closely approximate the PSR and have
the desirable property that $f(0)=0$. The efficiency function can
for example be defined as $f(\gamma_k)=f_s(\gamma_k)- f_s(0)$. In
almost all practical cases and for moderate to large values of $M$
(e.g. $M=100$), $f_s(0)$ is very small and, hence,
$f(\gamma_k)\simeq f_s(\gamma_k)$. In addition, for this
efficiency function we have $f(0)=0$. The plot of $f(\gamma_k)$
for the BPSK modulation and additive white Gaussian noise is given
in Fig. \ref{fig1} with $M=100$ (see \cite{Rod03} for a detailed
discussion of this efficiency function).

The exact expression for the efficiency function is not crucial.
Our analysis throughout this paper is valid for any efficiency
function that is increasing and S-shaped\footnote{An increasing
function is S-shaped if there is a point above which the function
is concave, and below which the function is convex.} with $f(0)=0$
and $f(+\infty)=1$, and has a continuous derivative. These
assumptions are valid in many practical systems. Throughout this
paper, we assume that all users have the same efficiency function.
Generalization to the case where the efficiency function is
dependent on $k$ is straightforward. Note that the throughput
$T_k$ in (\ref{eq4}) could also be replaced with the Shannon
capacity formula if the utility function in (\ref{eq4a}) is
appropriately modified to ensure that $u_k = 0$ when $p_k=0$.

\begin{figure}[t]
\centering
\includegraphics[width=3.5in]{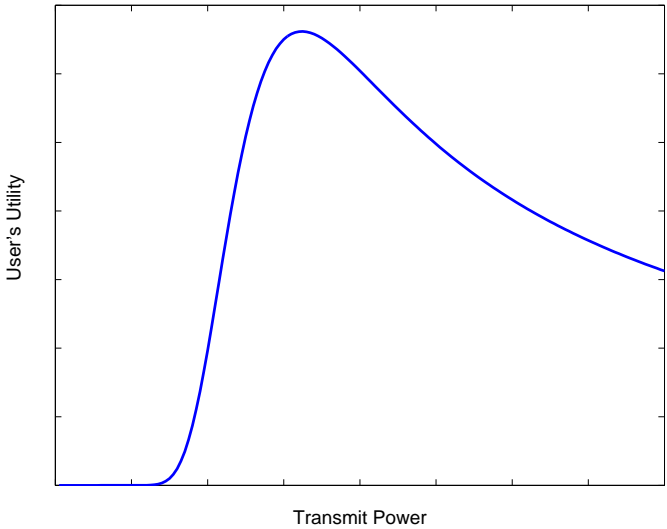}
\caption{User's Utility as a Function of Transmit Power for Fixed
Interference.} \label{fig2}
\end{figure}

Combining \eqref{eq4a} with \eqref{eq4}, and replacing the PSR
with the efficiency function, we can write the utility function of
the $k^{th}$ user as
\begin{equation}\label{eq5}
    u_k = \frac{L}{M} R_k \frac{f(\gamma_k)}{p_k} .
\end{equation}
This utility function, which has units of \emph{bits/Joule},
represents the total number of data bits that are delivered to the
destination without an error per joule of energy consumed. This
utility function captures very well the tradeoff between
throughput and battery life and is particularly suitable for
applications where saving power is more important than achieving a
high throughput. For the sake of simplicity, we assume that the
transmission rate is the same for all users, i.e.,
${R_1=\cdots=R_K=R}$. All the results obtained here can be easily
generalized to the case of unequal rates. Fig. \ref{fig2} shows
the shape of the utility function in (\ref{eq5}) as a function of
transmit power keeping other users' transmit powers fixed.

Power control is modelled as a non-cooperative game in which each
user tries to selfishly maximize its own utility. It is shown in
\cite{SaraydarThesis} that, when matched filters are used as the
uplink receivers, if user terminals are allowed to choose only
their transmit powers for maximizing their utilities, then there
exists an equilibrium point at which no user can improve its
utility given the power levels of other users (Nash equilibrium).
In this work, we extend this game-theoretic approach to study the
cross-layer design problem of joint multiuser detection and power
control. In particular, we propose a non-cooperative game in which
the users are allowed to choose their uplink receivers as well as
their transmit powers.

\section{System Model} \label{system model}

We consider the uplink of a DS-CDMA system with processing gain
$N$ (defined as the ratio of symbol duration to chip duration). We
assume that there are $K$ users in the network and focus on a
single cell. Thus, we assume that all $K$ user terminals transmit
to a receiver at a common concentration point, such as a cellular
base station or other network access point. For now, we assume
that each of the transmitters and the receiver has one antenna.
The signal received by the uplink receiver (after chip-matched
filtering) sampled at the chip rate over one symbol duration can
be expressed as
\begin{equation}\label{eq1}
    {\mathbf{r}} = \sum_{k=1}^{K} \sqrt{p_k} h_k \ b_k {\mathbf{s}}_k +
    {\mathbf{w}} ,
\end{equation}
where $p_k$, $h_k$, $b_k$ and ${\mathbf{s}}_k$ are the transmit
power, channel gain, transmitted bit and spreading sequence of the
$k^{th}$ user, respectively, and $\mathbf{w}$ is the noise vector
which is assumed to be Gaussian with mean $\mathbf{0}$ and
covariance $\sigma^2 \mathbf{I}$. We assume random spreading
sequences for all users, i.e., $ {\mathbf{s}}_k =
\frac{1}{\sqrt{N}}[v_1 ... v_N]^T$, where the $v_i$'s are
independent and identically distributed (i.i.d.) random variables
taking values \{$-1,+1$\} with equal probabilities.

Let us represent the linear uplink receiver of the $k^{th}$ user
by a coefficient vector, ${\mathbf{c}}_k$. The output of this
receiver can be written as
\begin{eqnarray}\label{eq2}
    y_k &=& {{\mathbf{c}}_k}^T {\mathbf{r}} \nonumber \\
        &=& \sqrt{p_k} h_k \ b_k {\mathbf{c}}_k^T {{\mathbf{s}}_k} + \sum_{j\neq k} \sqrt{p_j} h_j \ b_j {{\mathbf{c}}_k}^T {{\mathbf{s}}_j} +
    {\mathbf{c}}_k^T {\mathbf{w}}  .
\end{eqnarray}
Given (\ref{eq2}), the SIR of the $k^{th}$ user at the output of
its receiver is
\begin{equation}\label{eq3}
    \gamma_k = \frac{p_k h_k^2 ({\mathbf{c}}_k^T {\mathbf{s}}_k)^2} {\sigma^2 {\mathbf{c}}_k^T {\mathbf{c}}_k + \sum_{j\neq k} p_j h_j^2 ({\mathbf{c}}_k^T
    {\mathbf{s}}_j)^2}  .
\end{equation}

In all the previous work in this area, the receive filter is
assumed to be a simple matched filter and maximization of the
utility function is done over the transmit power only. In the
following section, we extend this approach by allowing the users
to choose their receivers in addition to their transmit powers. It
should be noted that although we focus on flat fading channels in
this paper, all of our analysis can be easily extended to
frequency-selective channels by appropriately defining an
effective spreading sequence for each user. In particular, the
effective spreading sequence for user $k$ is the response of the
frequency-selective channel to the transmitted spreading sequence
of user $k$.

\section{The Non-Cooperative Power Control Game} \label{NCPCG-SA}

Here, we propose a non-cooperative game in which each user seeks
to maximize its own utility by choosing its transmit power and the
receive filter coefficients. Let $G=~[{\mathcal{K}}, \{A_k \},
\{u_k \}]$ denote the proposed non-cooperative game where
${\mathcal{K}}=\{1, ... , K \}$, and $A_k=[0,P_{max}]\times
{\mathcal{R}}^N$ is the strategy set for the $k^{th}$ user. Here,
$P_{max}$ is the maximum allowed power for transmission. Each
strategy in $A_k$ can be written as ${\mathbf{a}}_k = (p_k ,
{\mathbf{c}}_k)$ where $p_k$ and ${\mathbf{c}}_k$ are the transmit
power and the receive filter coefficients, respectively, of user
$k$. Hence, the resulting non-cooperative game can be expressed as
the following maximization problem:
\begin{equation}\label{eq7b}
    \max_{{\mathbf{a}}_k} \ u_k =  \max_{p_k , \ {\mathbf{c}}_k} u_k (p_k, {\mathbf{c}}_k) \ \ \ \
    \textrm{for}
   \ \  k=1,...,K  .
\end{equation}

Assuming equal transmission rates for all users, (\ref{eq7b}) can
be expressed as
\begin{equation}\label{eq8}
    \max_{p_k , \ {\mathbf{c}}_k} \frac{f(\gamma_k (p_k, {\mathbf{c}}_k))}{p_k} \ \ \ \
    \textrm{for}
   \ \  k=1,...,K  ,
\end{equation}
where we have explicitly shown that $\gamma_k$ is a function of
$p_k$ and ${\mathbf{c}}_k$ as expressed in (\ref{eq3}). A Nash
equilibrium is a set of strategies such that no user can
unilaterally improve its own utility \cite{FudenbergTiroleBook91}.
We now state and prove the following proposition.\vspace{0.2cm}

\begin{proposition}
The Nash equilibrium for the non-cooperative game in (\ref{eq7b})
is given by $(p_k^*, {\mathbf{c}}_k^*)$ where ${\mathbf{c}}_k^*$
is the vector of MMSE receiver coefficients and $p_k^*=\min
(p_k^{MMSE}, P_{max})$. Here, $p_k^{MMSE}$ is the transmit power
that results in an SIR equal to $\gamma^*$, the solution to
$f(\gamma)=\gamma f'(\gamma)$, at the output of the MMSE receiver.
Furthermore, this equilibrium is unique (up to a scaling factor
for the MMSE filter coefficients).
\end{proposition}

\begin{thmproof}{Proof:}
We first show that at Nash equilibrium the receiver has to be the
MMSE receiver. Since the choice of receiver is independent of the
transmit power, we can write
\begin{eqnarray}\label{eq9}
    \max_{p_k , \ {\mathbf{c}}_k} \frac{f(\gamma_k (p_k, {\mathbf{c}}_k))}{p_k} &=& \max_{p_k} \frac{\max_{{\mathbf{c}}_k} f(\gamma_k (p_k,
    {\mathbf{c}}_k))}{p_k} \nonumber \\
     &=&  \max_{p_k} \frac{f(\max_{{\mathbf{c}}_k} \gamma_k (p_k,
    {\mathbf{c}}_k))}{p_k} \ ,
\end{eqnarray}
where the second equality is due to the fact that $f(\gamma)$ is
an increasing function of $\gamma$. It is well known that for any
given set of transmit powers, the MMSE receiver achieves the
maximum SIR among all linear receivers \cite{WangPoor}. Therefore,
the MMSE receiver achieves the maximum utility among all linear
receivers. For any output SIR, a user can always choose the MMSE
detector to achieve the desired SIR at a lower transmit power
compared to any other linear receiver. A lower transmit power
directly translates into a higher utility for the user. Therefore,
at Nash equilibrium (if it exists), the receiver must be the MMSE
receiver.

The SIR at the output of the MMSE receiver is given by
${\gamma_k^{MMSE}=p_k h_k^2 ({\mathbf{s}}_k^T {\mathbf{A}}_k^{-1}
{\mathbf{s}}_k)}$, where ${{\mathbf{A}}_k=\sum_{j\neq k} p_j h_j^2
{\mathbf{s}}_j {\mathbf{s}}_j^T+\sigma^2 \mathbf{I}}$. Since
${\frac{\partial \gamma_k^{MMSE}}{\partial
p_k}=\frac{\gamma_k^{MMSE}}{p_k}}$, maximizing the utility
function for each user is equivalent to finding $\gamma^*$ that is
the (positive) solution to ${f(\gamma_k)=\gamma_k \
f'(\gamma_k)}$\footnote{This is shown by taking the derivative of
$u_k$ with respect to $p_k$ and equating it to zero.}. If the
required power for achieving $\gamma^*$ is larger than $P_{max}$,
the utility function is maximized when $p_k=P_{max}$. Note that
$\gamma^*$ is independent of $k$ as long as all users have the
same efficiency function.

So far, we have shown that at Nash equilibrium (if it exists), the
receiver is the MMSE detector and each user's transmit power is
chosen to maximize the utility function with this set of filter
coefficients. Therefore, as in \cite{Saraydar02}, the existence of
the Nash equilibrium for the game in (\ref{eq7b}) can be shown via
the quasiconcavity of each user's utility function in its own
power\footnote{A function is quasiconcave if there exists a point
below which the function is non-decreasing, and above which the
function is non-increasing.}. For an S-shaped efficiency function,
with the MMSE detector as the receive filter,
$\frac{f(\gamma_k)}{p_k}$ is quasiconcave in $p_k$ and, hence, a
Nash equilibrium always exists.

Furthermore, for an S-shaped efficiency function,
${f(\gamma_k)=\gamma_k \ f'(\gamma_k)}$ has a unique solution,
$\gamma^*$, which is the (unique) maximizer of the utility
function \cite{Rod03b}. Because of the uniqueness of $\gamma^*$
and the one-to-one correspondence between the transmit power and
achieved SIR at the output of the MMSE receiver, the above Nash
equilibrium is unique.
\end{thmproof}\vspace{0.2cm}

The above equilibrium can be reached using the following iterative
algorithm. Given any set of users' transmit powers, the receiver
filter coefficients can be adjusted to the MMSE coefficients. Each
user can then adjust its transmit power to achieve $\gamma^*$ at
the output of the receiver. These steps can be repeated until
convergence is reached (see \cite{Yates95} for the proof of
convergence). It should be noted that $\gamma^*$ is the only SIR
value at which a line tangent to the curve describing $f(\gamma)$
passes through the origin (see \cite{Rod03b}). Throughout this
paper, we assume that $P_{max}$ is sufficiently large that
$\gamma^*$ can be achieved by all users.

In contrast to the traditional CDMA voice networks (e.g., IS-95)
where the target SIR is determined by the desired voice quality,
the common SIR here is determined by the utility function which in
turn is a function of the throughput which depends on the
modulation and coding schemes as well as the packet size.

\section{Comparison of Power Control Games for Linear Receivers} \label{comparison PC-SA}

In the previous section, we showed that the MMSE receiver achieves
the maximum utility among all linear receivers and hence is the
receiver chosen by users at the Nash equilibrium. In this section,
we fix the receiver type and allow users to choose their transmit
powers only. We focus on the matched filter, the decorrelator, and
the MMSE detector and obtain the Nash equilibrium for the
corresponding power control games. We show that irrespective of
the receiver type, the Nash equilibrium is an SIR-balancing
solution with the same target SIR for all receiver types. A
large-system analysis is then used to obtain closed-form
expressions for the utilities achieved at equilibria. This allows
us to compare the performance of these receivers in terms of the
number of bits transmitted per joule of energy consumed.

By picking a particular receiver, the power control game reduces
to
\begin{equation}\label{eq10}
    \max_{p_k} \frac{f(\gamma_k)}{p_k} \ \ \ \textrm{for} \ \ k=1,...,K .
\end{equation}
The relationship between the achieved SIR and transmit power
depends on the particular choice of the receiver. A necessary
condition for Nash equilibrium is that $\frac{\partial
u_k}{\partial p_k}=0$, i.e.,
\begin{equation}\label{eq11e}
    p_k\ \frac{\partial \gamma_k}{\partial p_k} f'(\gamma_k) - f(\gamma_k) = 0
    \ .
\end{equation}

We now examine this condition for the three detectors under
consideration.

For the conventional matched filter, we have
${\mathbf{c}}_k={\mathbf{s}}_k$ and, hence,
\begin{equation}\label{eq12}
    \gamma_k^{MF} = \frac{p_k h_k^2 } {\sigma^2 + \sum_{j\neq k} p_j h_j^2 ({\mathbf{s}}_k^T
    {\mathbf{s}}_j )^2 }\
    .
\end{equation}

For the decorrelator, we have ${\mathbf{C}}=[ {\mathbf{c}}_1\ ...\
{\mathbf{c}}_K ] = {\mathbf{S}} ( {\mathbf{S}} ^T {\mathbf{S}}
)^{-1}$ (for $K \leq N$), where ${\mathbf{S}}=[ {\mathbf{s}}_1\
...\ {\mathbf{s}}_K ]$. Hence,
\begin{equation}\label{eq13}
    \gamma_k^{DE} = \frac{p_k h_k^2 } {\sigma^2
    {\mathbf{c}}_k^T {\mathbf{c}}_k
    }\ .
\end{equation}

The filter coefficients for the MMSE receiver are given by
${\mathbf{c}}_k = \frac{\sqrt{p_k} h_k }{1+ p_k h_k^2
({\mathbf{s}}_k^T {\mathbf{A}}_k^{-1} {\mathbf{s}}_k)}
{\mathbf{A}}_k^{-1} {\mathbf{s}}_k $ (up to a scaling factor),
where ${\mathbf{A}}_k= \sum_{j\neq k} p_j h_j^2 {\mathbf{s}}_j
{\mathbf{s}}_j^T + \sigma^2 \mathbf{I}$. This results in
\begin{equation}\label{eq14}
    \gamma_k^{MMSE} = p_k h_k^2 ({\mathbf{s}}_k^T {\mathbf{A}}_k^{-1} {\mathbf{s}}_k)\
    .
\end{equation}

It is observed that for all three receivers, we have
\begin{equation}\label{eq15}
    \frac{\partial \gamma_k}{\partial p_k} = \frac{\gamma_k}{p_k} \ .
\end{equation}
Therefore, maximizing the utility function for each user is
equivalent to finding $\gamma^*$ that is the solution to
\begin{equation}\label{eq16}
    f(\gamma) = \gamma \ f'(\gamma)  \ .
\end{equation}

It is known that if $K$ and $N$ are large, the interference plus
noise term at the output of the matched filter can be approximated
as a Gaussian random variable \cite{Pursley77}. For the
decorrelator, since the multiple-access interference is removed
completely, the noise term at the output of the receiver is
Gaussian. In the case of the MMSE receiver, it has been shown that
the interference plus noise term at the output is well
approximated by a Gaussian random variable \cite{PoorVerdu97}.
Therefore, it is reasonable to assume that $f(\gamma)$ is the same
for these receivers. In addition, since the solution to
(\ref{eq16}) is the same for all users (provided that all users
have the same modulation and packet size), the solution to the
power control game is SIR-balanced with the same target SIR,
$\gamma^*$, independent of the choice of receiver.  Of course, the
amount of transmit power needed to achieve $\gamma^*$ is dependent
on the uplink receiver used and the channel gain (which will
depend, for example, on the distance between transmitter and
receiver). This means that while all the users achieve the same
throughput, their utilities will depend on their channel gains and
their receivers. In contrast to voice systems in which the target
SIR depends only on the desired quality of voice, here the target
SIR is dependent on the efficiency function and is influenced by
the modulation and packet size. It should be noted that as long as
the efficiency function is increasing and sigmoidal, $f(\gamma) =
\gamma \ f'(\gamma)$ has a unique solution.

Based on (\ref{eq12})--(\ref{eq14}), the amount of transmit power
required to achieve the target SIR, $\gamma^*$, will depend on the
random spreading sequence of each user. In order to obtain
quantitative results for the utility function corresponding to
each receiver, we appeal to a large system analysis similar to
that presented in \cite{TseHanly99}. We consider the asymptotic
case where $K, N \rightarrow \infty $ and $\frac{K}{N} \rightarrow
\alpha < \infty$. This allows us to write SIR expressions that are
independent of the spreading sequences of the users. It has been
shown in \cite{TseHanly99} that, for large systems, the SIR
expressions for the matched filter, the decorrelator and the MMSE
receiver are approximately given by

{\small{\begin{eqnarray}\label{eq17}
   &\gamma_k^{MF}&  = \frac{p_k h_k^2 } {\sigma^2 + \frac{1}{N}\sum_{j\neq k} p_j h_j^2 }\
     ,\\
\label{eq18}
 &\gamma_k^{DE}& = \frac{p_k h_k^2 (1-\alpha) } {\sigma^2 }\ \ \ \  {\textrm{for}} \ \ \alpha < 1 \
     ,\\
\textrm{and}
\label{eq19}
    &\gamma_k^{MMSE}& = \frac{p_k h_k^2 } {\sigma^2 + \frac{1}{N}\sum_{j\neq k} I(p_j h_j^2, p_k h_k^2, \gamma_k^{MMSE}) }\ ,
\end{eqnarray}}}
where $I(a,b,c)= \frac{ab}{b+ac}$.

It is clear that both $\gamma_k^{MF}$ and $\gamma_k^{DE}$ satisfy
$\frac{\partial \gamma_k}{\partial p_k} = \frac{\gamma_k}{p_k}$ .
It can also be verified that any $\gamma_k$ which satisfies
$\frac{\partial \gamma_k}{\partial p_k} = \frac{\gamma_k}{p_k}$ is
a solution to (\ref{eq19}). As a result, we claim that similar to
the previous section, finding the solution to power control in
large systems is equivalent to finding the solution of ${f(\gamma)
= \gamma \ f'(\gamma)}$, independent of the type of the receiver
used. Here again, the solution to this equation is independent of
$k$ which means that all users will seek to achieve the same SIR,
$\gamma^*$, at the output of the uplink receiver.

The minimum power solution for achieving $\gamma^*$ by all users
is given by the following equations for the three different
receivers (see \cite{TseHanly99}):
\begin{eqnarray} 
    &p_k^{MF}& = \frac{1}{h_k^2} \frac{\gamma^* \sigma^2}{1 - \alpha
    \gamma^*} \ \ \ \ \ {\textrm{for}} \ \ \alpha < \frac{1}{\gamma^*} \ \label{eq20-1} ,\\
   &p_k^{DE}& = \frac{1}{h_k^2} \frac{\gamma^* \sigma^2}{1 - \alpha} \ \
\ \ \ \ \ \ {\textrm{for}} \ \ \alpha < 1 \ \label{eq20-2}
    ,\\
\textrm{and}
    &p_k^{MMSE}& = \frac{1}{h_k^2} \frac{\gamma^* \sigma^2}{1 - \alpha
    \frac{\gamma^*}{1+\gamma^*}} \ \ {\textrm{for}} \ \ \alpha <
    1+\frac{1}{\gamma^*} \label{eq20-3}
    \ .
\end{eqnarray}
Combining (\ref{eq20-1})--(\ref{eq20-3}) with $u_k = \frac{L}{M} R
\frac{f(\gamma^*)}{p_k}$, we obtain
\begin{equation}\label{eq22b}
    u_k= \frac{ L R f(\gamma^*) h_k^2}{M \gamma^* \sigma^2}\ \Gamma \ ,
\end{equation}
where $\Gamma$ is dependent on the type of receiver. In
particular,
\begin{eqnarray} 
     &\Gamma^{MF}& = 1-\alpha \gamma^* \ \ \ \ \ \ \ \ \  {\textrm{for}} \ \ \alpha < \frac{1}{\gamma*}
    \ \label{eq23b-1}
    ,\\
    &\Gamma^{DE}& = 1-\alpha  \ \ \ \ \ \ \ \ \ \ \ \  {\textrm{for}} \ \ \alpha < 1
   \ \label{eq23b-2}
   ,\\
\textrm{and}
    &\Gamma^{MMSE}& = 1-\alpha \frac{\gamma^*}{1+\gamma^*} \ \ \ {\textrm{for}} \ \ \alpha < 1+\frac{1}{\gamma*} \ \label{eq23b-3} .
\end{eqnarray}
It can be seen that $u_k^{MMSE} \geq u_k^{DE}$ and $u_k^{MMSE}
\geq u_k^{MF}$ which confirms our earlier claim that the MMSE
receiver achieves the maximum utility among all linear receivers.
The utility achieved by the decorrelator is higher than that of
the matched filter except when $\gamma^* <1 (=0dB)$.

\section{Social Optimum}\label{social optimum}

The solution to the power control game is Pareto optimal if there
exists no other power allocation for which one or more users can
improve their utilities without reduction in the utilities of the
other users. It can be shown that the Nash equilibrium presented
in the previous section is not Pareto optimal. This means that it
is possible to improve the utility of one or more users without
hurting other users. On the other hand, it can be shown that the
solution to the following social problem gives the Pareto optimal
frontier (see \cite{SaraydarThesis}):
\begin{equation}\label{eq26}
    \max_{p_1,..., p_K} \sum_{k=1}^K \beta_k u_k (p_1,...,p_K) \ .
\end{equation}

Pareto optimal solutions are in general difficult to obtain. Here,
we consider the case of equal output SIRs among all users (i.e.,
SIR balancing). This ensures fairness among users in terms of
throughput and delay. We also assume that
${\beta_1=\cdots=\beta_K=1}$, which means we are interested in
maximizing the sum of users' utilities. Therefore, the
maximization in (\ref{eq26}) can be written as
\begin{equation}\label{eq27}
    \max_{p_1,..., p_K} f(\gamma) \sum_{k=1}^K \frac{1}{p_k} \ .
\end{equation}
Equal output SIRs among users is achieved with minimum power
consumption when the received powers are the same for all users,
i.e., $p_1 h_1^2 = p_2 h_2^2 = \dots = p_K h_K^2= q$, where
\begin{eqnarray}\label{eq28-1}
   &q^{MF}(\gamma)& =  \frac{\gamma \sigma^2}{1 - \alpha
    \gamma} \ \ \ \ \ {\textrm{for}} \ \ \alpha < \frac{1}{\gamma} \ ,\\
\label{eq28-2}
  &q^{DE}(\gamma)& =  \frac{\gamma \sigma^2}{1 - \alpha} \ \ \ \ \ \ \ {\textrm{for}} \ \ \alpha < 1 \
   ,\\
\textrm{and}
\label{eq28-3}
    &q^{MMSE}(\gamma)& = \frac{\gamma \sigma^2}{1 - \alpha
    \frac{\gamma}{1+\gamma}} \ \ {\textrm{for}} \ \ \alpha < 1+\frac{1}{\gamma}
    \ .
\end{eqnarray}
Therefore, the maximization in (\ref{eq27}) can equivalently be
expressed as
\begin{equation}\label{eq31}
    \max_{\gamma} \frac{f(\gamma)}{q(\gamma)} \sum_{k=1}^K h_k^2 \
    .
\end{equation}
The solution to (\ref{eq31}) must satisfy $\frac{\partial
}{\partial \gamma}(\frac{f(\gamma)}{q(\gamma)})=0$\ . Using this
fact, combined with (\ref{eq28-1})--(\ref{eq28-3}), gives us the
equations that must be satisfied by the solution to the
maximization problem in (\ref{eq31}) for the three linear
receivers:

{\small{\begin{eqnarray}\label{eq32-1}
     MF&:& \ f(\gamma)=\gamma (1-\alpha \gamma) f'(\gamma) \
    ,\\
\label{eq32-2}
     DE&:& \ f(\gamma)=\gamma f'(\gamma) \ ,\\
\textrm{and\ \ \ \ }
\label{eq32-3}
     MMSE&:& \ f(\gamma)=\gamma \Big{[} 1 - \frac{\alpha
    \gamma}{(1+\gamma)^2-\alpha \gamma^2} \Big{]}f'(\gamma)
    \ .
\end{eqnarray}}}
It should be noted that while the maximizations in (\ref{eq10})
and (\ref{eq31}) look similar, there is an important difference
between them. In (\ref{eq10}), the assumption is that there is no
cooperation among users. This means each user chooses its transmit
power independent of other users' powers. On the other hand,
(\ref{eq31}) assumes that users cooperate in choosing their
transmit powers. The consequence is that the relationship between
the user's SIR and transmit power is different from the
non-cooperative case.

\begin{figure}[t]
\centering \includegraphics[width=3.5in]{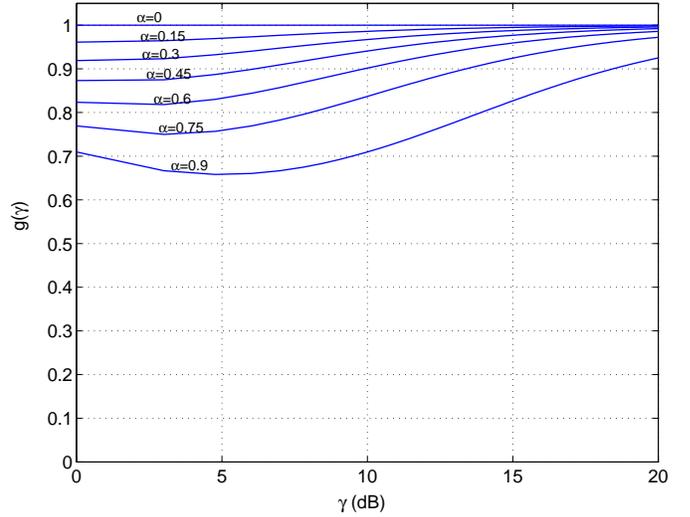} \caption{Plot
of $g(\gamma)=1-\frac{\alpha \gamma}{(1+\gamma)^2 - \alpha
\gamma^2}$ for Different Values of System Load.} \label{fig3}
\end{figure}

We see from (\ref{eq32-1})--(\ref{eq32-3}) that for the
decorrelator the Pareto-optimal solution is the same as the
solution obtained by the non-cooperative utility-maximizing
method. This is because the equilibrium transmit power of each
user is independent of other users' transmit powers (see
(\ref{eq13}) and (\ref{eq18})). Another observation is that as
shown in Fig. \ref{fig3}, for a large range of values of $\alpha$,
$1 - \frac{\alpha \gamma}{(1+\gamma)^2 - \alpha \gamma^2}\simeq
1$. This means that for the MMSE receiver, the target SIR for the
non-cooperative game, $\gamma^*$, is close to the target SIR for
the Pareto-optimal solution. This will be verified in Section
\ref{numerical results} using simulation.

\section{Extensions to Multi-antenna Systems}\label{PC-MA}

We now extend the analysis presented in the previous sections to
multi-antenna systems. In particular, we focus on the case of
receive diversity, i.e., multiple antennas at the uplink receiver.

We assume that each user terminal has one transmit antenna and
there are $m$ receive antennas at the uplink receiver. The
received signal (after chip-matched filtering and chip-rate
sampling) can be represented as an $N\times m$ matrix,
$\mathbf{R}$, where the $l^{th}$ column represents the $N$ chips
received at the $l^{th}$ antenna, i.e.,
\begin{equation}\label{eq1MA}
    {\mathbf{R}} = \sum_{k=1}^{K} \sqrt{p_k} \ b_k \ {\mathbf{s}}_k
    {\mathbf{h}}_k^T + \mathbf{W} \ ,
\end{equation}
where $p_k$, $b_k$ and ${\mathbf{s}}_k$ are the transmit power,
transmitted bit and spreading sequence of the $k^{th}$ user,
respectively. Here, ${\mathbf{h}}_k = [ h_{k1} \ ... \ h_{km} \
]^T$ represents the gain vector in which $h_{k1} , ... , h_{km}$
are the channel gains from the transmitter of the $k^{th}$ user to
the $m$ receive antennas and are assumed to be independent and
identically distributed. In the above equation, $\mathbf{W}$ is
the noise matrix. We assume that the noise is Gaussian and both
spatially and temporally white.

Let ${\mathbf{C}}_k$ be the $N\times m$ coefficient matrix for the
spatial-temporal filter of the $k^{th}$ user at the base station.
This filter performs linear spatial and temporal processing on the
received signal. The output of this receiver can be written as
\begin{equation}\label{eq2MA}
    y_k= tr({\mathbf{C}}_k^T \mathbf{R}) ,
\end{equation}
where $tr(\mathbf{A})$ is the \emph{trace} of $\mathbf{A}$.

Here, following an approach similar to that in Section
\ref{NCPCG-SA}, we propose a game in which users in the network
are allowed to choose their uplink linear spatial-temporal
receivers as well as their transmit powers. Hence, the resulting
non-cooperative game can be expressed as the following
maximization problem:
\begin{equation}\label{eq7MA}
    \max_{p_k , \ {\mathbf{C}}_k} u_k (p_k, {\mathbf{C}}_k) \ \ \ \
    {\textrm{for}}  \ \  k=1,...,K  .
\end{equation}

This maximization can equivalently be expressed as
\begin{equation}\label{eq8MA}
    \max_{p_k , \ {\mathbf{\bar{c}}}_k} u_k (p_k, {\mathbf{\bar{c}}}_k) \ \ \ \
    \textrm{for}
   \ \  k=1,...,K  ,
\end{equation}
where ${\mathbf{\bar{c}}}_k$ is a vector with $mN$ elements. It is
obtained by stacking the columns of ${\mathbf{C}}_k$ on top of
each other. To see this, notice that $y_k$ in (\ref{eq2MA}) can be
alternatively written as
\begin{equation}\label{eq9MA}
    y_k= {\mathbf{\bar{c}}}_k^T \mathbf{\bar{r}} \ ,
\end{equation}
where $\mathbf{\bar{r}}$ is a vector obtained by placing columns
of $\mathbf{R}$ on top of each other, i.e.,
\begin{equation}\label{eq10MA}
    {\mathbf{\bar{r}}} = \sum_{k=1}^{K} \sqrt{p_k} \ b_k {\mathbf{\bar{s}}}_k +
    {\mathbf{\bar{w}}} ,
\end{equation}
where  ${\mathbf{\bar{s}}}_k= [ h_{k1} {\mathbf{s}}_k^T \
 ... \ h_{km} {\mathbf{s}}_k^T \ ]^T$ is the \emph{effective
signature} and $\mathbf{\bar{w}}$ is the noise vector consisting
of columns of $\mathbf{W}$ stacked on top of each other.

The game expressed in (\ref{eq8MA}) is very similar to the one in
(\ref{eq7b}). As a result, all of our analysis for the single
antenna case can be carried over to the multi-antenna scenario.
Hence, we skip the analysis and state the main results:

\begin{itemize}
    \item The MMSE receiver, whose coefficients are given by
    \begin{equation}\label{eq39MA}
        {\mathbf{\bar{c}}}_k = \frac{\sqrt{p_k }}{1+ p_k ({\mathbf{\bar{s}}}_k^T {\mathbf{\bar{A}}}_k^{-1} {\mathbf{\bar{s}}}_k)} {\mathbf{\bar{A}}}_k^{-1} {\mathbf{\bar{s}}}_k \ ,
    \end{equation}
    achieves the maximum utility among all linear receivers.
    Here, ${\mathbf{\bar{A}}}_k= \sum_{j\neq k} p_j {\mathbf{\bar{s}}}_j {\mathbf{\bar{s}}}_j^T+ \sigma^2 \mathbf{I}$.
    \item Given the MMSE receiver coefficients, maximizing the utility function for each user is
    again equivalent to finding the solution $\gamma^*$ to $f(\gamma) = \gamma \ f'(\gamma)$.
    \item Nash equilibrium is reached when all user terminals use the MMSE detector for their uplink
    receivers and transmit at a power level that results in an SIR equal to
    $\gamma^*$ (SIR-balancing). This equilibrium is unique.
\end{itemize}

We now fix the receiver type and allow users to choose their
transmit powers only, as we did in Section \ref{comparison PC-SA}.
We again focus on the matched filter, the decorrelator and the
MMSE detector. We discuss the resulting Nash equilibria for these
three receivers and compare their performance using a large-system
analysis.

The matched filter is assumed to have perfect knowledge of the
channel gains of the desired user but knows only the statistics of
the fading levels of the interferers. It basically performs
despreading at each receive antenna and then applies maximal ratio
combining (MRC). The decorrelating detector is assumed to have
perfect knowledge of the channel gains for the desired user but no
knowledge about the interferers (except for their spreading
sequences). It applies a decorrelator at each receive antenna and
then performs maximal ratio combining. The MMSE detector is
assumed to have perfect knowledge of the channel gains of all
users. The filter coefficients for the MMSE receiver are given by
(\ref{eq39MA}).

It is straightforward to show that for all three receivers, we
have
\begin{equation}\label{eq22MA}
    \frac{\partial \gamma_k}{\partial p_k} = \frac{\gamma_k}{p_k} \ .
\end{equation}
Therefore, maximizing the utility function for each user is again
equivalent to finding $\gamma^*$ that is the solution to
$f(\gamma) = \gamma \ f'(\gamma)$. Using a large-system analysis
similar to the one presented in Section \ref{comparison PC-SA},
the achieved utility for user $k$ can be expressed as
\begin{equation}\label{eq25MA}
    u_k= \frac{L R f(\gamma^*) \bar{h}_k^2}{M \gamma^* \sigma^2}\
    \bar{\Gamma}\ ,
\end{equation}
where $\bar{\Gamma}$ depends on the receiver:
\begin{eqnarray}\label{eq26MA}
     &\bar{\Gamma}^{MF}& = 1-\bar{\alpha} \gamma^* \ \ \ \ \ \ \ \ \  {\textrm{for}} \ \ \bar{\alpha} < \frac{1}{\gamma*} \ , \\
\label{eq26MA-2}
    &\bar{\Gamma}^{DE}& = 1-\alpha  \ \ \ \ \ \ \ \ \ \ \ \  {\textrm{for}} \ \ \alpha < 1 \ ,  \\
\label{eq26MA-3}
    \textrm{and}  &\bar{\Gamma}^{MMSE}& = 1-\bar{\alpha} \frac{\gamma^*}{1+\gamma^*} \ \ \ {\textrm{for}} \ \ \bar{\alpha} < 1+\frac{1}{\gamma*}
    \ ,
\end{eqnarray}
with $\bar{\alpha}=\frac{\alpha}{m}$ and $\bar{h}_k^2 =
\sum_{l=1}^m h_{kl}^2$. It is observed that for the case of the
matched filter and the MMSE detector, using more antennas at the
receiver provides both power pooling (through $\bar{h}_k$) and
interference reduction (through $\bar{\alpha}$). This means that
the system behaves like a single-antenna system with processing
gain $mN$ and received power equal to the sum of the received
powers at the individual antennas. The decorrelator, on the other
hand, benefits only from power pooling and there is no pooling of
the degrees of freedom. This is because the decorrelating detector
has no knowledge about the channel gains for the interferers.
Therefore, each interferer effectively occupies $m$ degrees of
freedom \cite{HanlyTse01}.

\section{Utility-Maximizing Admission Control} \label{admission
control}

We have used a large-system analysis to derive explicit
expressions for the utilities achieved at Nash equilibrium for the
matched filter, the decorrelator, and the MMSE detector. We now
pose admission control as a maximization problem in which the load
in the network (i.e., $\alpha$) is chosen such that the total
utility in the network (per degree of freedom) is maximized:

\begin{equation}\label{eq62}
    \alpha^* = \arg \max_{\alpha} \frac{1}{N} \sum_{k=1}^{K} u_k   \ \ \ .
\end{equation}

Given (\ref{eq22b}), as $K,N \rightarrow \infty$, we can use the
law of large numbers to write
\begin{equation}\label{eq63}
        \alpha^* = \arg \max_{\alpha} \ \alpha \frac{L R f(\gamma^*)}{M \gamma^* \sigma^2} \
        \Gamma \
        {\mathbb{E}}\{h^2\}
        \ ,
\end{equation}
or equivalently
\begin{equation}\label{eq64}
    \alpha^* = \arg \max_{\alpha} \ \alpha \ \Gamma \    .
\end{equation}
To find $\alpha^*$, we set $\frac{\partial}{
\partial \alpha} (\alpha \Gamma) =0$ and solve for $\alpha$. It is easy to show that $\alpha^*$ is the solution to $\Gamma =
\frac{1}{2}$. This is also true for the Pareto-optimal solution
discussed in Section \ref{social optimum}. Given
(\ref{eq23b-1})--(\ref{eq23b-3}), we have
\begin{eqnarray}
   \alpha_{MF}^* &=& \frac{1}{2 \gamma^*} \ , \\
   \alpha_{DE}^* &=& \frac{1}{2} \ , \\
  \textrm{and} \ \ \alpha_{MMSE}^* &=& \frac{1}{2} + \frac{1}{2 \gamma^*} \ .
\end{eqnarray}

Following an argument similar to that above, it can be shown that
for the case of multiple receive antennas, the total utility per
degree of freedom is maximized when $\alpha$ is chosen to be the
solution to $\bar{\Gamma}=\frac{1}{2}$. Notice that using this
utility-maximizing admission control scheme, the number of
admitted users for the MMSE receiver is greater than or equal to
the total number of admitted users for the matched filter and
decorrelator combined (depending on the number of antennas
employed at the uplink receiver).

\section{Numerical Results} \label{numerical results}
In this section, we present numerical results for the analysis
presented in the previous sections. We consider the uplink of a
DS-CDMA system. We assume that each packet contains 100 bits of
information and no overhead (i.e., $L=M=100$). The transmission
rate, $R$, is $100Kbps$ and the thermal noise power, $\sigma^2$,
is $5\times 10^{-16}Watts$. The processing gain is 100 to satisfy
the large system assumption. A useful example for the efficiency
function is $f(\gamma)= (1- e^{-\gamma})^M$. This serves as an
approximation to the PSR that is very reasonable for moderate to
large values of $M$. We use this efficiency for our simulations.
Using this, with $M=100$, the solution to (\ref{eq16}) is
$\gamma^*=6.48 = 8.1dB$.

\begin{figure}[t]
\centering
\includegraphics[width=3.5in]{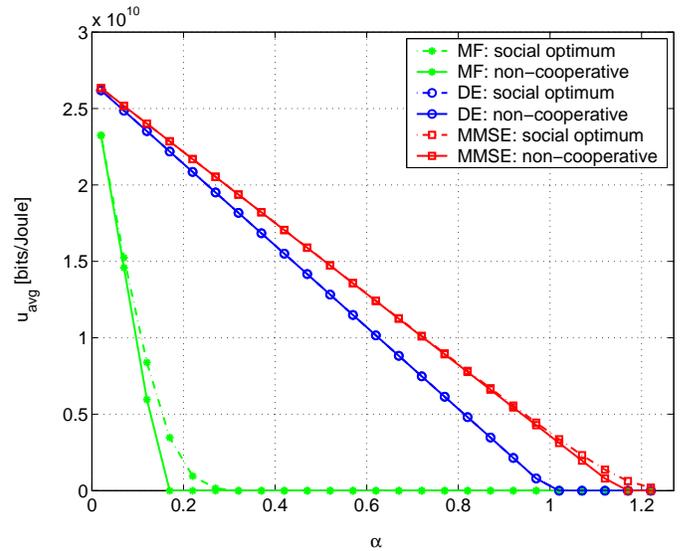}
\caption{Average Utility vs Load for the Matched Filter (MF), the
Decorrelator (DE) and the MMSE Receiver (Single Receive Antenna).}
\label{fig4}
\end{figure}

\begin{figure}[t]
\centering \includegraphics[width=3.5in]{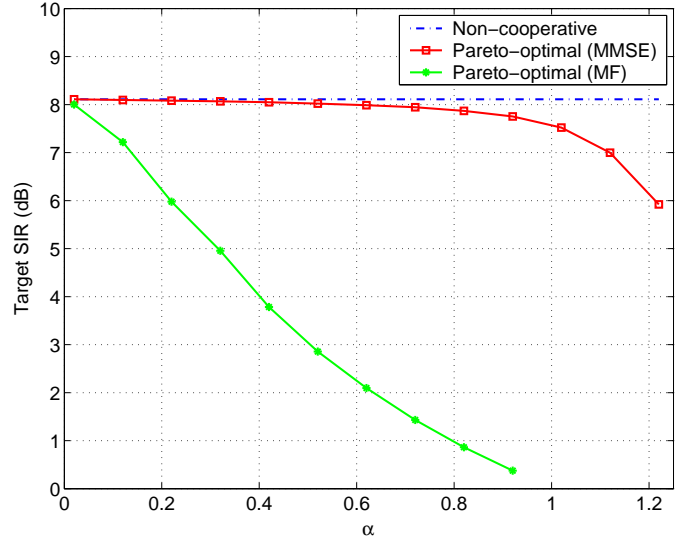}
\caption{Comparison of the Target SIRs for the Non-cooperative and
Pareto-optimal Solutions.} \label{fig5}
\end{figure}

We first look at the case of one receive antenna.  The channel
gains are assumed to have the Rayleigh distribution with mean
equal to $\frac{0.3}{d^2}$, where $d$ is the distance of the user
from the uplink receiver. Fig. \ref{fig4} shows the average
utility of a user as a function of the system load for the matched
filter, decorrelator and MMSE receivers. The user is assumed to be
$100$ meters away from the uplink receiver. The averaging is done
over 5000 channel realizations. The solid and dashed lines
correspond to the non-cooperative and Pareto-optimal solutions,
respectively. It is seen from the figure that the utility improves
considerably when the matched filter is replaced by a multiuser
detector. Also, the system capacity (i.e., the maximum number of
users that can be accommodated by the system) is larger for the
multiuser receivers as compared with the matched filter. As
expected, the MMSE receiver achieves the highest utility. While
the difference between the non-cooperative approach and the
Pareto-optimal solution is significant for the matched filter, the
solutions are identical for the decorrelator and are quite close
to each other for the MMSE receiver. It is seen that for the
matched filter, as the system load increases, the gap between the
non-cooperative and Pareto-optimal solutions becomes larger. This
is also true for the MMSE receiver (although much less
noticeably). Fig. \ref{fig5} compares the target SIR of the
non-cooperative solutions with the target SIRs of the
Pareto-optimal solutions for the matched filter and the MMSE
detector. It is seen that for the MMSE receiver, the target SIR
for the Pareto-optimal solution is very close to the target SIR
for the non-cooperative approach.

\begin{figure}[t]
\centering \includegraphics[width=3.5in]{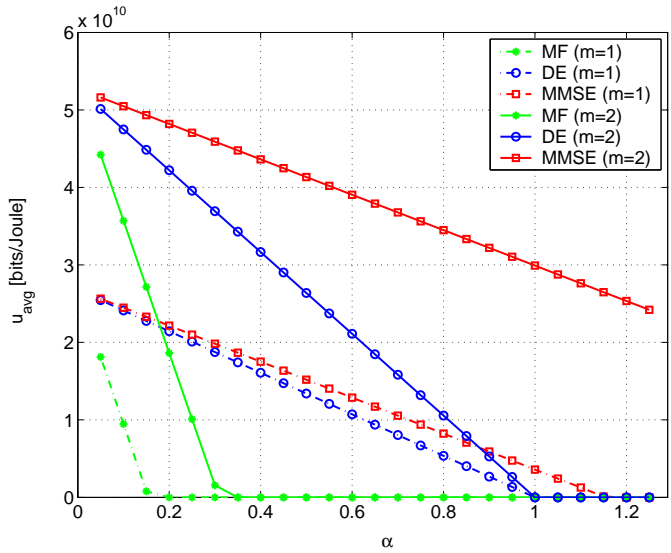}
\caption{Average Utility vs Load for the Matched Filter (MF), the
Decorrelator (DE) and the MMSE Receiver with One and Two Receive
Antennas}. \label{fig6}
\end{figure}

\begin{figure}[t]
\centering \includegraphics[width=3.5in]{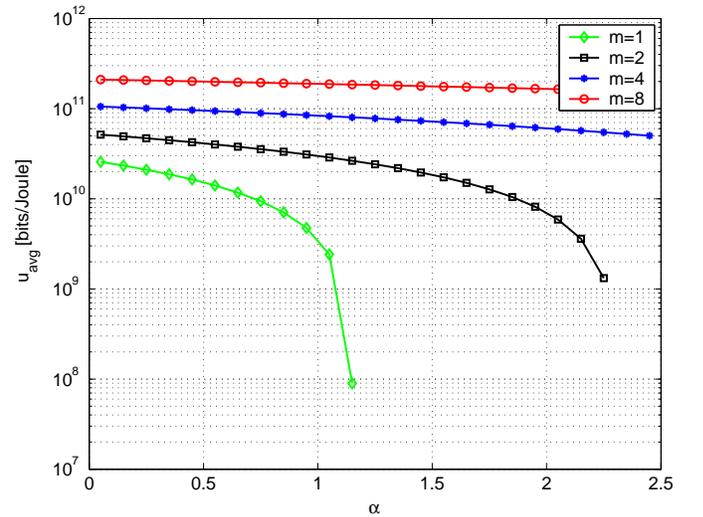}
\caption{Average Utility vs Load for the MMSE Receiver with $m$
Receive Antennas}. \label{fig7}
\end{figure}

Fig. \ref{fig6} shows the average utility as a function of the
system load for one and two receive antennas. The user is $100$
meters away from the uplink receiver and the channel gains are
assumed to be i.i.d. with a Rayleigh distribution having a mean
equal to $\frac{0.3}{d^2}$. The averaging is done over 5000
realizations of the channel gains. For each realization, we use
(\ref{eq25MA}) to calculate the user's utility. The figure shows
the achieved utilities for the matched filter, the decorrelator
and the MMSE receiver. The dashed lines correspond to $m=1$
(single receive antenna) and the solid lines represent the case of
$m=2$ (two receive antennas). Significant improvements in user
utility and system capacity are observed when two receive antennas
are used compared to the single antenna case. As expected, the
improvement is more significant for the matched filter and the
MMSE receiver as compared with the decorrelating detector. This is
because the matched filter and the MMSE receiver benefit from both
power pooling and interference reduction whereas the decorrelating
detector benefits only from power pooling. Fig. \ref{fig7} shows
the average utility versus system load for the MMSE receiver for
the cases of one, two, four and eight receive antennas. It is seen
that adding more antennas at the uplink receiver results in
considerable gains in the achieved utility as well as system
capacity.

\begin{figure}[t]
\centering \includegraphics[width=3.5in]{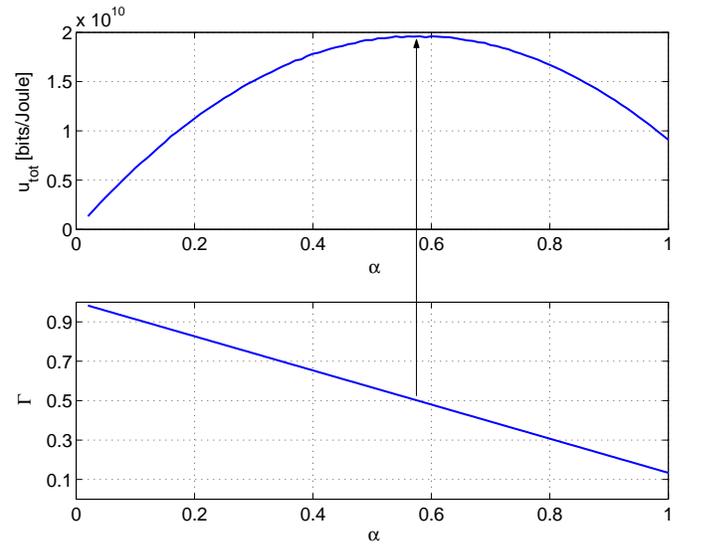} \caption{Plots
of Total Utility and $\Gamma$ vs Load for the MMSE Receiver
(Single Receive Antenna).} \label{fig8}
\end{figure}

We now look at the utility-maximizing admission control. We
consider the MMSE receiver and plot the total utility as a
function of system load. For each value of $\alpha$, we distribute
the users in the cell and calculate each user's utility according
to (\ref{eq22b}). We then calculate the total utility and repeat
this over 10 000 realizations of the users' locations. Fig.
\ref{fig8} shows the plot of average total utility versus system
load. We have also plotted $\Gamma$ as a function of $\alpha$. As
expected, the total utility is maximized when
$\Gamma=\frac{1}{2}$. This corresponds to a system load of 58\%.

\section{Conclusions} \label{conclusions}

In this work, we have examined the cross-layer design problem of
joint multiuser detection and power control in the uplink of CDMA
systems using a game-theoretic approach. A non-cooperative game is
proposed in which users are allowed to choose not only their
transmit powers but also their uplink receivers to maximize their
utilities. Focusing on linear receivers, we have shown that there
is a unique Nash equilibrium for the proposed game. The
equilibrium is achieved when all users pick the MMSE detector as
their uplink receivers and choose their transmit powers such that
their output SIRs are all equal to $\gamma^*$. We have further
shown that the Nash equilibrium remains an SIR-balancing solution
when we replace the MMSE receiver with a matched filter or a
decorrelating detector (or any other linear receiver). The target
SIR is affected by the modulation as well as the packet size but
is independent of the receiver. However, the utilities achieved at
equilibria do depend on the receiver. Using a large-system
analysis, we have obtained explicit expressions for the utilities
achieved at equilibrium by the matched filter, the decorrelator
and the MMSE detector, and compared their performance in terms of
number of bits transmitted per joule of energy consumed.
Significant improvements in achieved utilities and system capacity
have been observed when multiuser detectors are used in place of
the conventional matched filter. We have also discussed the
optimum cooperative solution and compared its performance with
that of the non-cooperative approach. It has been shown that the
difference in performance is not significant especially for the
decorrelator and the MMSE receiver.

We have also extended our approach to systems with multiple
receive antennas. Conclusions similar to those for the single
antenna case have been made. We have shown that considerable gains
in achieved utilities and system capacity are obtained when
multiple antennas are employed at the uplink receiver. These
gains, which are due to power pooling and interference reduction,
are quantified in terms of number of bits transmitted per joule of
energy. A utility-maximizing admission control scheme has also
been proposed. We have shown that using the proposed scheme, the
total number of admitted users for the MMSE receiver is greater
than or equal to the total number of admitted users for the
matched filter and decorrelator, depending on the number of
receive antennas. This work has provided a unified game-theoretic
formulation for studying power control and receiver design in
DS-CDMA networks.

\vspace{-0.5cm}
\begin{biography}{Farhad Meshkati} received the B.A.Sc. degree in engineering science
(electrical option) and the M.A.Sc. degree in electrical
engineering in 1999 and 2001, respectively, both from the
University of Toronto, Canada. He is currently pursuing the Ph.D.
degree in electrical engineering at Princeton University,
Princeton, NJ. He spent the summer of 2004 with the Corporate R\&D
Systems group at Qualcomm,~Inc., in San Diego, CA. His research
interests are in wireless communication, especially radio resource
management.
\end{biography}
\begin{biography}{H. Vincent Poor}  (S'72, M'77, SM'82, F'87) received
the Ph.D. degree in EECS from Princeton University in 1977. From
1977 until 1990, he was on the faculty of the University of
Illinois at Urbana-Champaign. Since 1990 he has been on the
faculty at Princeton, where he is the George Van Ness Lothrop
Professor in Engineering. Dr. Poor's research interests are in the
areas of statistical signal processing and its applications in
wireless networks and related fields. Among his publications in
these areas is the recent book \emph{Wireless Networks: Multiuser
Detection in Cross-Layer Design} (Springer: New York, NY, 2005).

Dr. Poor is a member of the National Academy of Engineering and
the American Academy of Arts and Sciences, and is a Fellow of the
Institute of Mathematical Statistics, the Optical Society of
America, and other organizations.  In 1990, he served as President
of the IEEE Information Theory Society, and in 1991-92 he was a
member of the IEEE Board of Directors.  He is currently serving as
the Editor-in-Chief of the \emph{IEEE Transactions on Information
Theory}. Recent recognition of his work includes the Joint Paper
Award of the IEEE Communications and Information Theory Societies
(2001), the NSF Director's Award for Distinguished Teaching
Scholars (2002), a Guggenheim Fellowship (2002-03), and the IEEE
Education Medal (2005).
\end{biography}
\begin{biography}{Stuart C. Schwartz} received the B.S. and M.S. degrees from M.I.T. in 1961 and the
Ph.D.  from the University of Michigan in 1966. While at M.I.T. he
was associated with the Naval Supersonic Laboratory and the
Instrumentation Laboratory (now the Draper Laboratories).  During
the year 1961-62 he was at the Jet Propulsion Laboratory in
Pasadena, California, working on problems in orbit estimation and
telemetry.  During the academic year 1980-81, he was a member of
the technical staff at the Radio Research Laboratory, Bell
Telephone Laboratories, Crawford Hill, NJ, working in the area of
mobile telephony.

He is currently a Professor of Electrical Engineering at Princeton
University.  He was chair of the department during the period
1985-1994, and served as Associate Dean for the School of
Engineering during the period July 1977-June 1980.  During the
academic year 1972-73, he was a John S. Guggenheim Fellow and
Visiting Associate Professor at the department of Electrical
Engineering, Technion, Haifa, Israel. He has also held visiting
academic appointments at Dartmouth, University of California,
Berkeley, and the Image Sciences Laboratory, ETH, Zurich.  His
principal research interests are in statistical communication
theory, signal and image processing.
\end{biography}
\newpage
\begin{biography}{Narayan B. Mandayam} (S'90-M'95-SM'00) received the B.Tech. (Hons.) degree in 1989 from
the Indian Institute of Technology (IIT), Kharagpur, India, and
the M.S. and Ph.D. degrees in 1991 and 1994, respectively, from
Rice University, Houston, TX, all in electrical engineering.

From 1994 to 1996, he was a Research Associate at the Wireless
Information Network Laboratory (WINLAB), Rutgers University, New
Brunswick, NJ. In September 1996, he joined the faculty of the ECE
Department at Rutgers, where he became Associate Professor in 2001
and Professor in 2003. Currently, he also serves as Associate
Director at WINLAB. He was a Visiting Faculty Fellow in the
Department of Electrical Engineering, Princeton University,
Princeton, NJ, in Fall 2002, and Visiting Faculty at the Indian
Institute of Science, Bangalore, India, in Spring 2003. His
research interests are in various aspects of wireless data
transmission, including system modeling and performance, signal
processing, and radio resource management, with emphasis on
open-access techniques for spectrum sharing. He is a coauthor
(with C. Comaniciu and H. V. Poor) of the book Wireless Networks:
Multiuser Detection in Cross-Layer Design, New York, Springer,
2005.

Dr. Mandayam is a recipient of the Institute Silver Medal from IIT
Kharagpur in 1989 and the U.S. National Science Foundation CAREER
Award in 1998. He has served as an Editor for IEEE COMMUNICATION
LETTERS from 1999-2002, and currently serves as an Editor for the
IEEE TRANSACTIONS ON WIRELESS COMMUNICATIONS.
\end{biography}

\end{document}